\documentstyle[12pt,epsfig]{article}
\newcommand{\bea}{\begin{eqnarray}}
\newcommand{\eea}{\end{eqnarray}}
\newcommand{\be}{\begin{equation}}
\newcommand{\ee}{\end{equation}}
\newcommand{\nn}{\nonumber}

\begin{document}
\large
\vspace{6cm}
\begin{center}
{\Large\bf On nature of scalar and tensor mesons from the analysis of processes $\pi\pi\to\pi\pi,K\overline{K},\eta\eta$}\\
\vspace{0.5cm}

{\bf Yu.S.~Surovtsev}$^a$, {\bf R.~Kami\'nski}$^b$,
{\bf D.~Krupa}$^c$ and {\bf M.~Nagy}$^c$ \\
\vspace{0.2cm}
$^a$ Bogoliubov Laboratory of Theoretical Physics, Joint Institute
for Nuclear Research, Dubna 141 980, Russia\\
$^b$ Institute of Nuclear Physics, Polish Academy of Sciences,
PL 31 342 Cracow, Poland\\
$^c$ Institute of Physics, Slovak Academy of Sciences, D\'ubravsk\'a
cesta 9, 845 11 Bratislava, Slovakia

\end{center}

\vspace{0.1cm}

{\large
\begin{center}
{\bf Abstract}\\
\end{center}
Analysis of the isoscalar $S$- and $D$-waves of processes 
$\pi\pi\to\pi\pi,K\overline{K},\eta\eta$ is carried out aimed at studying the status and QCD nature of scalar and tensor mesons below 2 GeV and 2.3 GeV, respectively. Assignment of these mesons to lower scalar and tensor nonets is proposed.\\

\vspace{.5cm}

{\bf Outline:}
\begin{itemize}
\item Motivation
\item Three-coupled-channel formalism
\item Model-independent analysis of isoscalar-scalar sector
\item Lower scalar nonets
\item Model analysis of isoscalar-tensor sector
\item Discussion and conclusions
\end{itemize}

\section{Motivation}
The study of spectrum of low-lying hadrons and their properties is
very important for investigation of the confinement problem
and elaboration of nonperturbative methods of QCD. Especially it
concerns scalar mesons. E.g., two different nonperturbative
methods of QCD (QCD sum rules and unquenched lattice
calculations) give essentially distinct results for the lightest
scalar glueball.\\
\underline{QCD sum rules:}g
a lightest scalar meson with a mass below 900 MeV is rather narrow and
non-$q{\bar q}$ state (glueball) -- see, {\it e.g.},
\cite{Elias,Narison,Kisslinger}.\\
\underline{Lattice simulations:} The lowest mass state of a pure glue
should be the $0^{++}$ with a mass of 1670$\pm$20 MeV \cite{Michael}.

An assignment of the discovered scalar mesonic states to
quark-model configurations is problematic up to now.

It is very important to have model-independent information on
investigated states and on their QCD nature. It can be obtained only on
the basis of the first principles (analyticity and unitarity)
directly applied to analysis of experimental data. That approach permits
us to introduce no theoretical prejudice into extracted parameters of
resonances. We have already
proposed such method \cite{KMS-nc96}. Here we have applied it to combined
analysis of experimental data on the processes
$\pi\pi\to\pi\pi,K\overline{K},\eta\eta$ in the channel with the quantum
numbers $I^GJ^{PC}=0^+0^{++}$. In the considered 3-channel case, it is
turned out to be possible to use a method of the uniformizing variable
which takes into account the Riemann surface structure.
Considering the obtained disposition of resonance poles on the Riemann
surface, obtained coupling constants with channels and resonance masses,
we draw definite conclusions about nature of the investigated states.

Further we analyze the same processes in the channel with the quantum
numbers $I^GJ^{PC}=0^+2^{++}$ for the study of the $f_2$ mesons below
2.3 GeV. In this sector, from thirteen discussed resonances, the nine
ones ($f_2(1430)$, $f_2(1565)$, $f_2(1640)$, $f_2(1810)$, $f_2(1910)$,
$f_2(2000)$, $f_2(2020)$, $f_2(2150)$, $f_2(2220)$) must be confirmed in
various experiments and analyses \cite{PDG-04}. Recently in the combined analysis of
$p\overline{p}\to\pi\pi,\eta\eta,\eta\eta^\prime$, five resonances --
$f_2(1920)$, $f_2(2000)$, $f_2(2020)$, $f_2(2240)$ and $f_2(2300)$ --
have been obtained, one of which ($f_2(2000)$) is a candidate for the
glueball \cite{Anisovich}.

In the tensor sector, in addition to the indicated three channels, we
consider explicitly also the channel $(2\pi)(2\pi)$. In the 4-channel
case it is impossible to apply the uniformizing-variable method with
using a simple variable.
Therefore, the resonance poles are generated by the some 4-channel
Breit-Wigner forms with taking into account a Blatt-Weisskopf barrier
factor \cite{Blatt-Weisskopf} conditioned by the resonance spins.

\section{Three-coupled-channel formalism}
The $S$-matrix for the 3-channel case is determined on
the 8-sheeted Riemann surface. The elements $S_{\alpha\beta}$,
where $\alpha,\beta=1(\pi\pi), 2(K\overline{K}),3(\eta\eta)$,
have the right-hand cuts along the real axis of the $s$ complex
plane, starting with $4m_\pi^2$, $4m_K^2$, and $4m_\eta^2$, and
the left-hand cuts. The sheets of surface are numbered according
to the signs of analytic continuations of the channel momenta
$$~k_1=(s/4-m_\pi^2)^{1/2},~~~k_2=(s/4-m_K^2)^{1/2},~~~
k_3=(s/4-m_\eta^2)^{1/2}$$ as follows:

\begin{center}
\begin{tabular}{|c|c|c|c|c|c|c|c|c|} \hline
{} & I & II & III & IV & V & VI & VII & VIII \\ \hline
{$\mbox{Im}k_1$} & $+$ & $-$ & $-$ & $+$ & $+$ & $-$ & $-$ & $+$ \\
{$\mbox{Im}k_2$} & $+$ & $+$ & $-$ & $-$ & $-$ & $-$ & $+$ & $+$\\
{$\mbox{Im}k_3$} & $+$ & $+$ & $+$ & $+$ & $-$ & $-$ & $-$ & $-$\\ \hline
\end{tabular}
\end{center}
The resonance representations on the Riemann surface are obtained
with the help of formulas from Ref.~\cite{KMS-nc96},
expressing analytic continuations of the matrix elements to unphysical
sheets in terms of those on sheet I that have only zeros (beyond the
real axis) corresponding to resonances, at least, around the physical
region.
Then, starting from resonance zeros on sheet I, we obtain 7 types of
resonances corresponding to 7 possible situations when there are
resonance zeros on sheet I only in ({\bf a}) $S_{11}$; ({\bf b}) $S_{22}$;
({\bf c}) $S_{33}$; ({\bf d}) $S_{11}$ and $S_{22}$; ({\bf e}) $S_{22}$
and $S_{33}$; ({\bf f}) $S_{11}$ and $S_{33}$; and ({\bf g}) $S_{11}$,
$S_{22}$, and $S_{33}$.
E.g., the arrangement of poles corresponding to a resonance of type
({\bf g}) is: each sheet II, IV, and VIII contains a pair of conjugate
poles at the points that are zeros on sheet I; each sheet III, V, and
VII contains two pairs of conjugate poles; and sheet VI contains three
pairs of poles. A resonance of every type is represented by a pair of
complex-conjugate clusters (of poles and zeros on the Riemann surface).
Representation of multichannel resonances by the pole clusters gives
a main model-independent effect of resonances.
The cluster kind is related to nature of the state. The resonance
coupled relatively more strongly to the $\pi\pi$ channel than to the
$K\overline{K}$ and $\eta\eta$ ones is described by the cluster of type
({\bf a}); the resonance with dominant $s{\bar s}$ component, by the
cluster of type ({\bf e}); the glueball, by the ({\bf g}) cluster.
Note that at usual representation of multichannel resonances by the simple
Breit-Wigner forms, cases ({\bf d}),({\bf e}), ({\bf f}) and ({\bf g})
practically are lost.

We can distinguish, in a model-independent way, a bound state of
colourless particles ({\it e.g.}, $K\overline{K}$ molecule) and a
$q{\bar q}$ bound state \cite{KMS-nc96,MP-93}. 

For the combined analysis of data we use the Le Couteur-Newton relations
\cite{CL}:
$$
S_{11}=\frac{d(-k_1,k_2,k_3)}{d(k_1,k_2,k_3)},~~
S_{22}=\frac{d(k_1,-k_2,k_3)}{d(k_1,k_2,k_3)},~~
S_{33}=\frac{d(k_1,k_2,-k_3)}{d(k_1,k_2,k_3)},\nn
$$
\begin{equation}
S_{11}S_{22}-S_{12}^2=\frac{d(-k_1,-k_2,k_3)}{d(k_1,k_2,k_3)},~~
S_{11}S_{33}-S_{13}^2=\frac{d(-k_1,k_2,-k_3)}{d(k_1,k_2,k_3)}.
\end{equation}

The Jost matrix determinant $d(k_1,k_2,k_3)$ is the real analytic
function with the only square-root branch-points at $k_i=0$.

In the model-independent approach that is based only on the first
principles (analyticity-microcausality and unitarity) and is free from
dynamical assumptions, we use the {\it mathematical} fact that a local
behaviour of analytic functions determined on the Riemann surface is
governed by the nearest singularities on all corresponding sheets.
To take into account the branch points, we must find proper uniformizing
variable. However, it is impossible to map the 8-sheeted Riemann surface
onto a plane with the help of a simple function.
Therefore, we neglect the influence of the $\pi\pi$ threshold (however,
unitarity on the $\pi\pi$ cut is taken into account). This approximation
means the consideration of the semi-sheets of the Riemann surface nearest to
the physical region. The uniformizing variable can be chosen as
\cite{KMS-nc96}
\begin{equation}
w=\frac{k_2+k_3}{\sqrt{m_\eta^2-m_K^2}}.
\end{equation}
It maps our model of the 8-sheeted Riemann surface onto the $w$-plane.

On the $w$-plane, the Le Couteur-Newton relations are somewhat modified:
$$
S_{11}=\frac{d^* (-w^*)}{d(w)},~~
S_{22}=\frac{d(-w^{-1})}{d(w)},~~ S_{33}=\frac{d(w^{-1})}{d(w)},
$$
\begin{equation}
S_{11}S_{22}-S_{12}^2=\frac{d^*({w^*}^{-1})}{d(w)},~~
S_{11}S_{33}-S_{13}^2=\frac{d^*(-{w^*}^{-1})}{d(w)}.~~
\end{equation}
The $d$-function is taken as
$$
d=d_B d_{res}
$$
where the resonance part is
\begin{equation}
d_{res}(w)=w^{-\frac{M}{2}}\prod_{r=1}^{M}(w+w_{r}^*)
\end{equation}
with $M$ the number of resonance zeros.
$d_B$, describing the background, is
\begin{equation}
d_B=\mbox{exp}[-i\sum_{n=1}^{3}\frac{k_n}{m_n}(\alpha_n+i\beta_n)],
\end{equation}
where
\begin{equation}
\alpha_n=a_{n1}+a_{n\sigma}\frac{s-s_\sigma}{s_\sigma}\theta(s-
s_\sigma)+ a_{nv}\frac{s-s_v}{s_v}\theta(s-s_v),
\end{equation}
\begin{equation}
\beta_n=b_{n1}+b_{n\sigma}\frac{s-s_\sigma}{s_\sigma}\theta(s-s_\sigma)+
b_{nv}\frac{s-s_v}{s_v}\theta(s-s_v).
\end{equation}
The second terms in $\alpha_n$ and $\beta_n$ take into account effectively possible channels below roughly
1400 MeV (mainly $\sigma\sigma$-channel); the third terms, many opening channels
in the range of about 1.5 GeV ($\eta\eta^{\prime},\rho\rho,\omega\omega$);
$s_v$ is their combined threshold.
Moreover, the $\pi\pi$ background is taken to be elastic up to the
$K\overline{K}$ threshold.

\section{Model-independent analysis of isoscalar-scalar sector}
We analyzed the data on three processes $\pi\pi\to\pi\pi,K\overline{K},\eta\eta$
in the channel with the vacuum quantum numbers. As the data, we use the results
of phase analyses which are given for phase shifts of the amplitudes
$\delta_{ab}$ and for moduli of the $S$-matrix elements $\eta_{ab}=|S_{ab}|$
(a,b=1-$\pi\pi$,~2-$K\overline{K}$,~3-$\eta\eta$):
\begin{equation}
S_{aa}=\eta_{aa}e^{2i\delta_{aa}},~~~~~S_{ab}=\eta_{ab}e^{i\delta_{ab}}.
\end{equation}
If below the $\eta\eta$-threshold there is the 2-channel unitarity, then the
relations
\begin{equation}
\eta_{11}=\eta_{22}, ~~ \eta_{12}=(1-{\eta_{11}}^2)^{1/2},~~
\delta_{12}=\delta_{11}+\delta_{22}
\end{equation}
are fulfilled in this energy region.

For the $\pi\pi$ scattering, the data from the threshold to 1.89 GeV are
taken from Ref. \cite{Hyams}; below 1 GeV, from many works 
\cite{Zylbersztejn}.

For $\pi\pi\to K\overline{K}$, practically all the accessible data are used
\cite{Wetzel}.
For the process $\pi\pi\to\eta\eta$, here we exploited data for the quantity
$|S_{13}|^2$ from the threshold to 1.72 GeV \cite{Binon}.

We considered the case with all five resonances discussed below 1.9 GeV.
From a variety of variants of the resonance representations by possible
pole-clusters, the analysis selects the following one to be most relevant
-- when the $f_0 (600)$ is described by the cluster of type ({\bf a});
$f_0 (1370)$, type ({\bf c}); $f_0 (1500)$, type ({\bf g}); $f_0 (1710)$,
type ({\bf b}); the $f_0 (980)$ is represented only by the pole on sheet
II and shifted pole on sheet III. Description of the resonances of types
({\bf a}), ({\bf b}) and ({\bf c}) can be related to the Breit-Wigner forms.
To reduce the number of adjusted parameters, we make it here, except for
the $f_0 (980)$).

We obtain a satisfactory description: for the $\pi\pi$-scattering from
about 0.4~GeV to 1.89~GeV ($\chi^2/\mbox{ndf}=202.111/(165-34)\approx 1.54$);
for the process $\pi\pi\to K\overline{K}$, from the threshold to
about 1.6~GeV ($\chi^2/\mbox{ndf}=161.912/(120-33)\approx 1.86$); for the $|S_{13}|^2$
data of the reaction $\pi\pi\to\eta\eta$, from the threshold to 1.72~GeV
($\chi^2/\mbox{ndf}\approx 0.992$). The total $\chi^2/\mbox{ndf}$ for all
three processes is $379.893/(301-42)\approx1.46$. \\The background parameters
are
~$a_{11}=0.183$, $a_{1\sigma}=0.0252$, $a_{1v}=0.0155$,
$b_{11}=0$, $b_{1\sigma}=-0.0089$, $b_{1v}=0.04336$,
$a_{21}=-0.6973$, $a_{2\sigma}=-1.427$, $a_{2v}=-5.935$,
$b_{21}=0.0447$, $b_{2\sigma}=0$, $b_{2v}=7.044$,
$b_{31}=0.6346$, $b_{3\sigma}=0.3336$, $b_{2v}=0$;
$s_\sigma=1.638~{\rm GeV}^2$, $s_v=2.084~{\rm GeV}^2$.

On figures 1-5, we demonstrate energy dependences of phase
shifts and moduli of the matrix elements of processes
$\pi\pi\to \pi\pi,K\overline{K},\eta\eta$ compared with the
experimental data.

\begin{figure}[ht]
\centering
\epsfxsize=17cm\epsffile{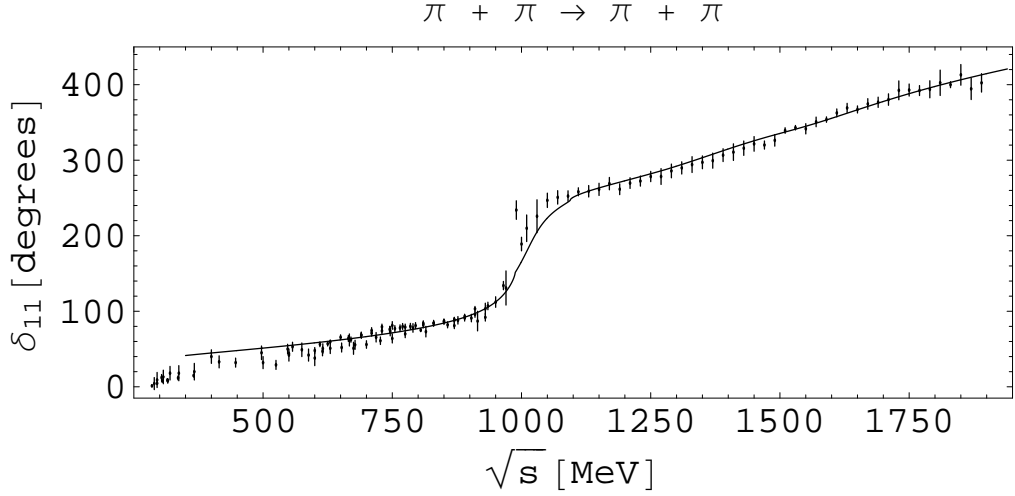}
\caption{The phase shift of the $\pi\pi$-scattering $S$-wave amplitude.}
\end{figure}

\vspace*{0.5cm}

\begin{figure}[ht]
\centering
\epsfxsize=17cm\epsffile{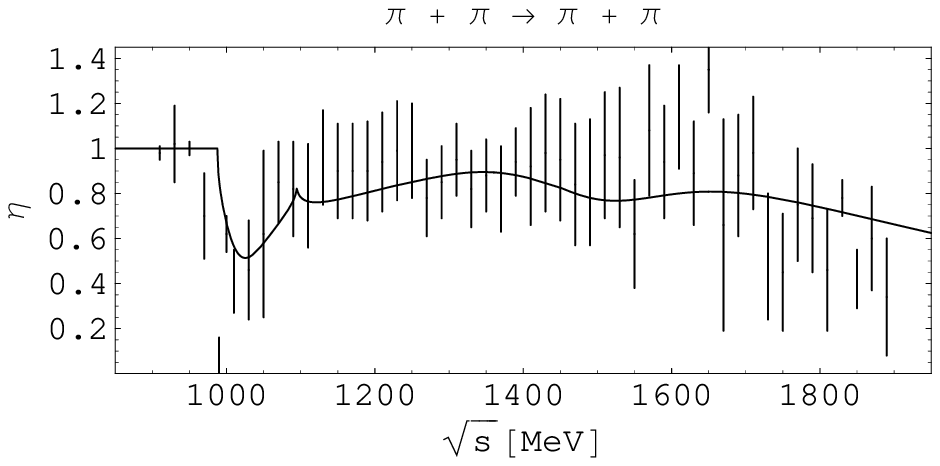}
\caption{The module of the $\pi\pi$-scattering $S$-wave matrix element.}
\end{figure}

\vspace*{-0.1cm}
\begin{figure}[ht]
\centering
\epsfxsize=17cm\epsffile{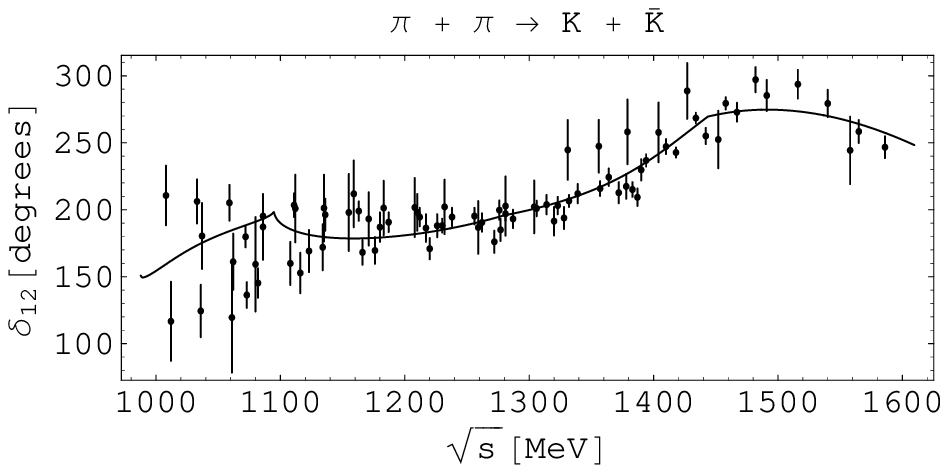}
\caption{The phase shift of the $\pi\pi\to K\overline{K}$ $S$-wave matrix
element.}
\end{figure}

\vspace*{-0.2cm}

\begin{figure}[ht]
\centering
\epsfxsize=17cm\epsffile{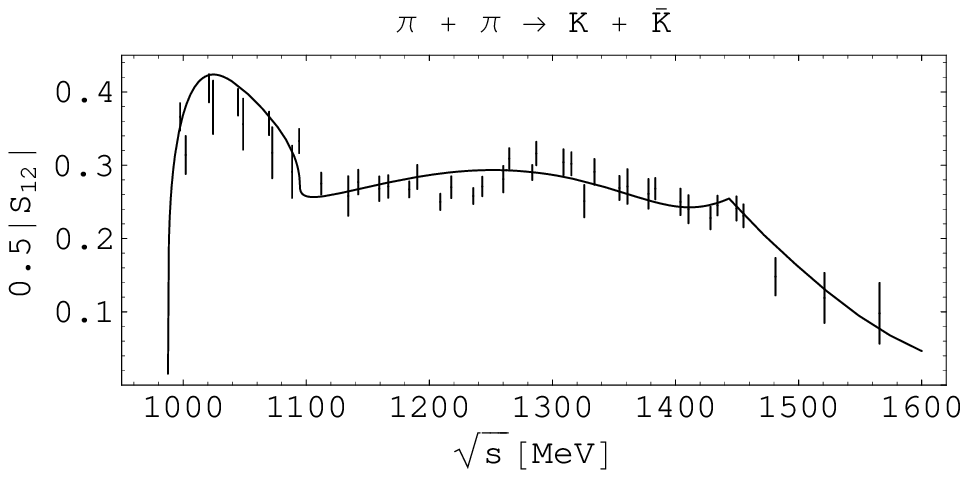}
\caption{The module of the $\pi\pi\to K\overline{K}$ $S$-wave matrix element.}
\end{figure}

\vspace*{-0.2cm}

\begin{figure}[ht]
\centering
\epsfxsize=17cm\epsffile{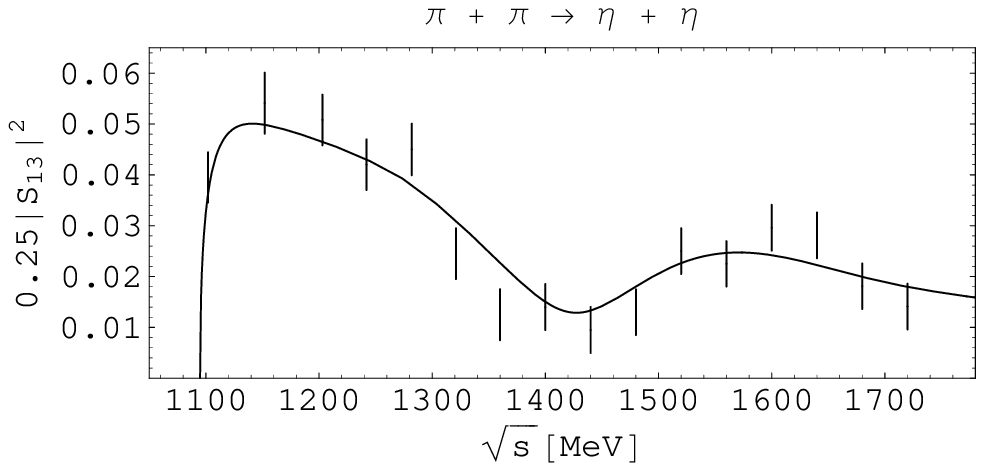}
\caption{The squared module of the $\pi\pi\to\eta\eta$ $S$-wave matrix element.}
\end{figure}

Let us indicate in Table 1 the obtained pole clusters for resonances on the
complex energy plane $\sqrt{s}$, poles on sheets IV, VI, VIII and V,
corresponding to the $f_0 (1500)$, are of the 2nd and 3rd order,
respectively (this is an approximation).
\hspace*{-1cm}
{\normalsize
\begin{table}[ht] \centering\caption{Pole clusters for the $f_0$-resonances.}
\vskip0.3truecm
\begin{tabular}{|c|c|c|c|c|c|c|c|c|} \hline \multicolumn{2}{|c|}{Sheet} &
II & III & IV & V & VI & VII & VIII \\ \hline {$f_0 (600)$} & {${\rm E}_r$} &
678$\pm$14 & 688$\pm$16 & {} & {} & 628$\pm$17 & 618$\pm$15 & {} \\ {} &
{$\Gamma_r$} & 608$\pm$22 & 608$\pm$9 & {} &{} & 608$\pm$28 & 608$\pm$26 &
{}\\ \hline {$f_0(980)$} & {${\rm E}_r$} & 1016$\pm$5 & 986$\pm$18 & {} & {} & {}
& {} & {}\\ {} & {$\Gamma_r$} & 32$\pm$8 & 59$\pm$16 & {} & {} & {} & {} &
{} \\ \hline {$f_0 (1370)$} & {${\rm E}_r$}
& {} & {} & {} & 1400$\pm$21 & 1400$\pm$20 & {1400$\pm$20} & {1400$\pm$20} \\
{} & {$\Gamma_r$}
& {} & {} & {} & 89$\pm$13 & 71$\pm$15 & {45$\pm$6} & {27$\pm$9} \\
\hline {$f_0 (1500)$} & {${\rm E}_r$}
& 1505$\pm$22 & 1480$\pm$30 & 1505$\pm$20 & 1500$\pm$20 & 1493$\pm$27
& 1488$\pm$25 & 1505$\pm$20 \\ {} & {$\Gamma_r$}
& 360$\pm$23 & 140$\pm$21 & 240$\pm$30 & 139$\pm$21 & 194$\pm$27 & 88$\pm$15
& 360$\pm$30 \\ \hline
{$f_0 (1710)$} & {${\rm E}_r$}
& {} & 1704$\pm$18 & 1704$\pm$21 & 1704$\pm$32 & 1704$\pm$30 & {} & {} \\
{} & {$\Gamma_r$}
& {} & 95$\pm$14 & 105$\pm$17 & 325$\pm$26 & 325$\pm$45 & {} & {} \\ \hline  \end{tabular}
\label{tab:cluster34} \end{table}}

Note a surprising result obtained for the $f_0(980)$ state.
It turns out that this state lies slightly above the $K\overline{K}$
threshold and is described by a pole on sheet II and by a shifted pole on
sheet III under the $\eta\eta$ threshold without an accompaniment of the
corresponding poles on sheets VI and VII, as it was expected for standard
clusters. This corresponds to the description of the $\eta\eta$ bound
state.

For now, we did not calculate coupling constants in the 3-channel
approach.
Therefore, for subsequent conclusions, let us mention the results for
coupling constants from our previous 2-channel analysis (Table 2)\cite{SKN-EPJA}:
$g_{1r}$ is the coupling constant of resonance "r" with the $\pi\pi$-system; $g_{2r}$, with $K\overline{K}$.
{\large
\begin{table}[htb]\centering\caption{Coupling constants of the $f_0$-resonances
from the 2-channel analysis.}
\begin{tabular}{|c|c|c|c|c|} \hline {}  &
$f_0(665)$ & $f_0(980)$ &
$f_0(1370)$ & $f_0(1500)$\\ \hline $g_{1r}$, GeV
& ~$0.652\pm 0.065$~ &
~$0.167 \pm 0.05$~~ & ~$0.116 \pm 0.03$~ &
~$0.657 \pm 0.113$~\\ \hline
$g_{2r}$, GeV & ~$0.724\pm0.1$~~~~ &
~$0.445\pm0.031$~ & ~~$0.99\pm0.05$~ &
~$0.666\pm0.15$~~\\ \hline \end{tabular}
\label{tab:constants} \end{table}}
We see that the $f_0(980)$ and the ${f_0}(1370)$ are coupled essentially
more strongly to the $K\overline{K}$ system than to the $\pi\pi$ one,
{\it i.e.}, they have a dominant $s{\bar s}$ component. The $f_0(1500)$ has
the approximately equal coupling constants with the $\pi\pi$ and
$K\overline{K}$, which apparently could point to its dominant
glueball component.
In the 2-channel case, $f_0 (1710)$ is represented by the cluster
corresponding to a state with the dominant $s{\bar s}$ component.

Our 3-channel conclusions on the basis of resonance cluster types
generally confirm the ones drawn in the 2-channel analysis
(besides the above surprising conclusion about the $f_0(980)$ nature).

Masses and widths of these states should be calculated from the pole
positions. If to take the resonance part of amplitude as
\begin{equation}
T^{res}=\sqrt{s}\Gamma_{el}/(m_{res}^2-s-i\sqrt{s}\Gamma_{tot}),
\end{equation}
we obtain for masses and total widths the following values (in MeV):

{\begin{center}{for~~ $f_0 (600)$,~~ 868~~ and~~ 1212;\\
for~~ $f_0 (980)$,~~ 1015.5~~ and~~ 64;\\
for $f_0 (1370)$,~~ 1407.5~~ and~~ 344;\\
for ~$f_0 (1500)$,~~~ 1546~~ and~~ 716;\\
for $f_0 (1710)$,~~ 1709.6~~ and~~ 276.}\end{center}}

\section{Lower scalar nonets}
It is known that an assignment of the scalar mesonic states to
quark-model configurations is problematic up to now, although there is a
number of interesting conjectures \cite{Lanik}-\cite{Volkov-Yudichev}.
Let us also (on the basis of obtained results) propose a following
assignment of scalar mesons below 1.9 GeV to lower nonets. First of all,
we exclude from this consideration the $f_0(980)$ as the $\eta\eta$ bound
state. Then we propose to include to the lowest nonet the isovector
$a_0(980)$, the isodoublet $K_0^*(905)$ (or $\kappa(800)$), and two
isoscalars $f_0(600)$ and $f_0(1370)$ as mixtures of the eighth component
of octet and the SU(3) singlet. Note that we consider the $K_0^*(905)$
(or $\kappa$) which one has observed at analysing the $K-\pi$ scattering 
\cite{Ishida,Bugg}, extracted from reaction $K^-p\to K^-\pi^+n$, 
and at studying the decay $D^+\to K^-\pi^+\pi^+$ \cite{Aitala}. Then the
Gell-Mann--Okubo formula
\begin{equation}
3m_{f_8}=4m_{K_0^*}-m_{a_0} \nn
\label{GM-O}
\end{equation}
gives $m_{f_8}=0.88$ GeV. Our result for the $\sigma$-meson mass is
$m_\sigma\approx0.868\pm0.02$ GeV (if $m_\kappa=0.8$, $m_{f_8}\approx0.73$).

In the relation for masses of nonet
\begin{equation}
m_\sigma+m_{f_0(1370)}=2m_{K_0^*(905))}, \nn
\label{8-0-1/2}
\end{equation}
the left-hand side is about 26 \% bigger than the right-hand one if to
take our mass values.

The next nonet (maybe, of radial excitations) could be formed of the
isovector $a_0(1450)$, the isodoublet $K_0^*(1430-1450)$, and of the
$f_0(1500)$ and $f_0(1710)$ as mixtures of the eighth component of octet and the SU(3) singlet, the $f_0(1500)$ being mixed with a glueball which is dominant in this state.
From the Gell-Mann--Okubo formula we obtain $m_{f_8}\approx1.45$ GeV.
In second formula
\begin{equation}
m_{f_0(1500)}+m_{f_0(1710)}=2m_{K^*(1450)},\nn
\label{r:8-0-1/2}
\end{equation}
the left-hand side is about 12 \% bigger than the right-hand one.

Though the Gell-Mann--Okubo formula is fulfilled for both nonets rather
satisfactorily, however, the breaking of 2nd relation (especially for the
lowest nonet) tells us that the $\sigma-f_0(1370)$ and $f_0(1500)-f_0(1710)$
systems get additional contributions absent in the $K_0^*(905)$ and
$K_0^*(1450)$, respectively.

Note that the 3-channel analysis indicates on a non-simple picture of mixing
the states $f_0(1370)$ and $f_0 (1710)$ with the wide $f_0(1500)$ and $f_0(600)$.
The $f_0(1370)$ is coupled more strongly to the $\eta\eta$ channel than to the
$\pi\pi$ and $K\overline{K}$ ones; the $f_0 (1710)$, to the $K\overline{K}$ than
$\pi\pi$ and $\eta\eta$ ones, whereas if these states were the pure $s{\bar s}$
ones, they would be described by clusters of type ({\bf e}), and their coupling
constants with the $K\overline{K}$ and $\eta\eta$ channels would be congruent
numbers for each state.

\section{Model analysis of isoscalar-tensor sector}

We analyze the isoscalar D-waves of the processes
$\pi\pi\to\pi\pi,K\overline{K},\eta\eta$ in the 4-channel approach with the
explicit account of the channel $(2\pi)(2\pi)$ (i=4), too. The Jost matrix
determinant $d(k_1,k_2,k_3,k_4)$ is taken as
\begin{equation}
d=d_B d_{res}.
\end{equation}
The 4-channel Breit-Wigner form for the resonance part of the
$d$-function is taken in the form ($\rho_{rj}=2k_i/\sqrt{M_r^2-4m_j^2}$):
\begin{equation}
d_{res}(s)=\prod_{r}
\left[M_r^2-s-i\sum_{j=1}^4\rho_{rj}^5R_{rj}f_{rj}^2\right],
\end{equation}
where $f_{rj}^2/M_r$ is the partial width, $R_{rj}$ is a Blatt-Weisskopf barrier
factor \cite{Blatt-Weisskopf}
\begin{equation}
R_{rj}=\frac{9+\frac{3}{4}
(\sqrt{M_r^2-4m_j^2}~r_{rj})^2+\frac{1}{16}(\sqrt{M_r^2-4m_j^2}~r_{rj})^4}
{9+\frac{3}{4}(\sqrt{s-4m_j^2}~r_{rj})^2+\frac{1}{16}(\sqrt{s-4m_j^2}~r_{rj})^4}
\end{equation}
with radii of 0.955 Fermi for all resonances in all channels, except for
$f_2(1270)$, $f_2^{\prime}(1525)$ and $f_2(1950)$ for which they are: for
$f_2(1270)$,~ 1.496, 0.704 and 0.604 Fermi respectively in channels $\pi\pi$,
$K\overline{K}$ and $\eta\eta$ ,
for $f_2^{\prime}(1525)$, ~0.576 and 0.584 Fermi in channels $K\overline{K}$ and
$\eta\eta$, and for $f_2(1950)$, ~0.178 Fermi in channel $K\overline{K}$.

The background is parameterized by
\begin{equation}
 d_B=\mbox{exp}\left[-i\sum_{n=1}^{3}\left(\frac{2k_n}{\sqrt{s}}\right)^5(a_n+
 ib_n)\right].
\end{equation}
To take into account an influence of opening channels in the range of
$\sim$ 1.5 GeV ($\eta\eta^{\prime},\rho\rho,\omega\omega$), $a_1$
and $b_n~(n=1,2,3)$ are taken in the form:
\begin{equation}
a_1=\alpha_{11}+\frac{s-4m_K^2}{s}\alpha_{12}\theta(s-4m_K^2)+
\frac{s-s_v}{s}\alpha_{10}\theta(s-s_v)),
\end{equation}
\begin{equation}
b_n=\beta_n+\frac{s-s_v}{s}\gamma_n\theta(s-s_v),
\end{equation}
where $s_v\approx2.274$ GeV$^2$ is the combined threshold of channels $\eta\eta^{\prime},~
\rho\rho,~\omega\omega$.

The data for the $\pi\pi$ scattering are taken from an energy-independent analysis
by B.~Hyams et al. \cite{Hyams}.
The data for the processes $\pi\pi\to K\overline{K},\eta\eta$ are taken from
works \cite{Lindenbaum}. We obtained ten resonances
$f_2(1270)$, $f_2(1430)$, $f_2^{\prime}(1525)$, $f_2(1580)$, $f_2(1730)$,
$f_2(1810)$, $f_2(1950)$, $f_2(2000)$, $f_2(2240)$ and $f_2(2410)$. We will not
discuss the last resonance because there are not practically data. We need it for
satisfying unitarity.

On figures 6-9, we demonstrate results from our fitting
to data (in the Argand plot units).
\begin{figure}[ht]
\centering
\epsfxsize=17cm\epsffile{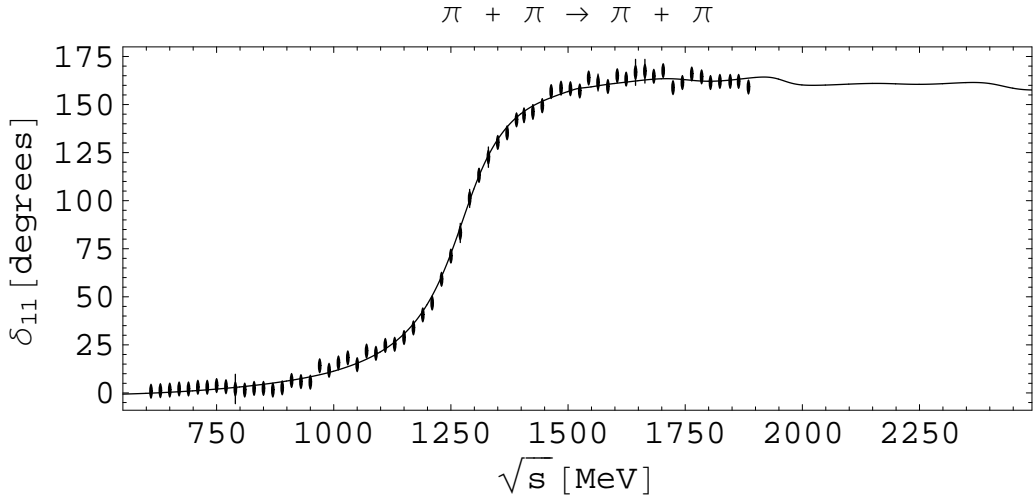}
\caption{The phase shift of the $\pi\pi$-scattering $D$-wave amplitude.}
\end{figure}
\vspace{1cm}

\begin{figure}[ht]
\centering
\epsfxsize=17cm\epsffile{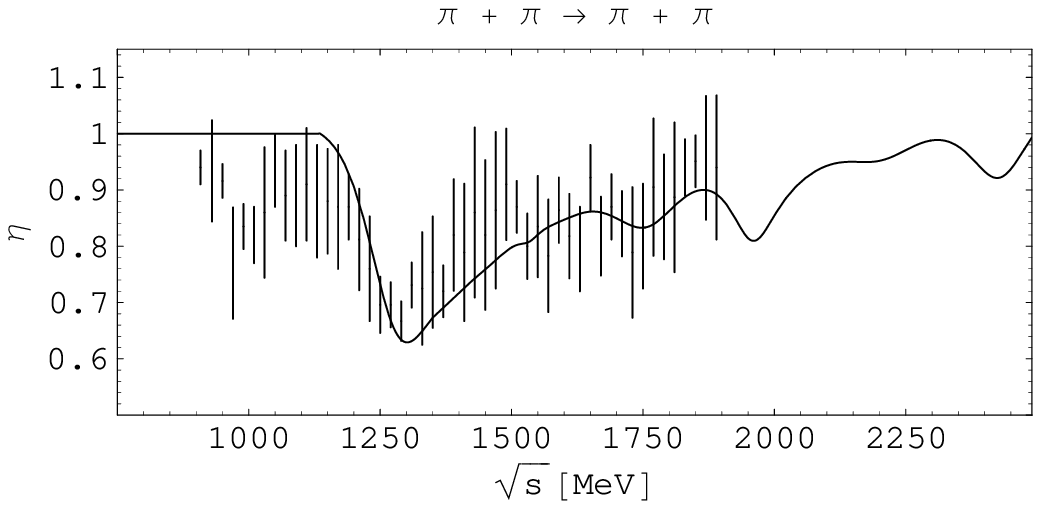}
\caption{The module of the $\pi\pi$-scattering $D$-wave matrix element.}
\end{figure}

\begin{figure}[ht]
\centering
\epsfxsize=17cm\epsffile{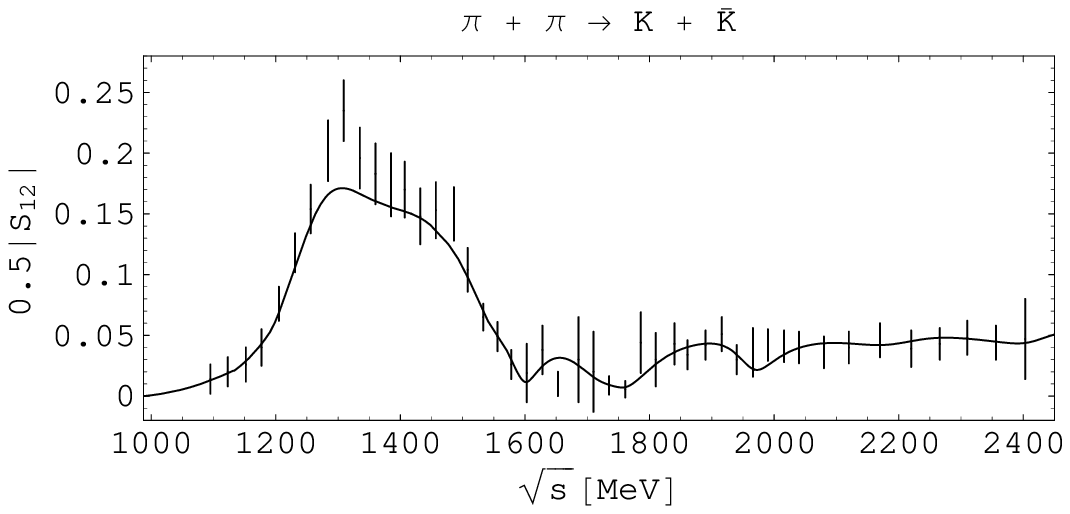}
\caption{The squared module of the $\pi\pi\to K\overline{K}$ $D$-wave matrix
element.}
\end{figure}
\vspace{1cm}

\begin{figure}[ht]
\centering
\epsfxsize=17cm\epsffile{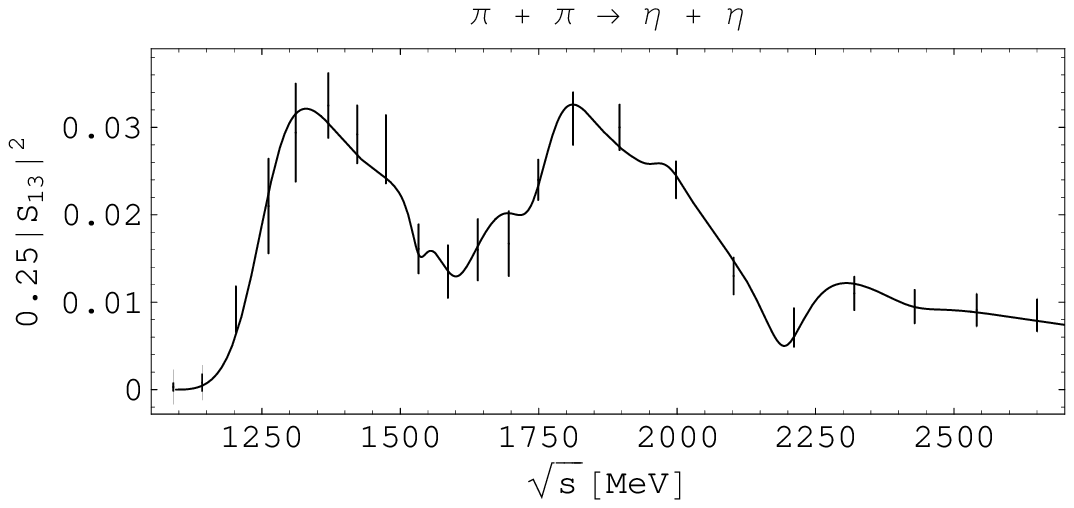}
\caption{The squared module of $\pi\pi\to\eta\eta$ $D$-wave matrix element.}
\end{figure}

We obtain a reasonable description (the total
$\chi^2/\mbox{ndf}=162.577/(168-66)\approx 1.59$) with the values of parameters
of $f_2$-resonances shown in Table 3.

{\normalsize
\begin{table}[ht]
\centering\caption{The $f_2$-resonance parameters (all in the MeV units).}
\vskip0.3truecm
\begin{tabular}{|c|c|c|c|c|c|c|} \hline
{State} & ~$M$~ & $f_{r1}$ & $f_{r2}$ & $f_{r3}$ & $f_{r4}$ &  $\Gamma_{tot}$ \\
\hline
{$f_2(1270)$} & 1275.1$\pm$1.8 & 470.9$\pm$5.4 & 201.5$\pm$11.4 & 89.5$\pm$4.76 &
22.6$\pm$4.6 & $>$212\\
{$f_2(1430)$} & 1450.8$\pm$18.7 & 128.3$\pm$45.9 & 562.3$\pm$142 & 32.7$\pm$18.4 &
8.2$\pm$65 & $>$230\\
{$f_2^{\prime}(1525)$} & 1535$\pm$8.6 & 28.6$\pm$8.3 & 253.8$\pm$78 & 92.7$\pm$11.5 &
41.4$\pm$160 & $>$76\\
{$f_2(1565)$} & 1601.4$\pm$27.5 & 75.5$\pm$19.4 & 315$\pm$48.6 & 388.9$\pm$27.7 &
127$\pm$199 & $>$170\\
{$f_2(1730)$} & 1724.4$\pm$5.7 & 78.8$\pm$43 & 289.5$\pm$62.4 & 460.3$\pm$54.6 &
107.6$\pm$76.7 & $>$181\\
{$f_2(1810)$} & 1766.5$\pm$15.3 & 129.5$\pm$14.4 & 259$\pm$30.7 & 469.7$\pm$22.5 &
90.3$\pm$90 & $>$177\\
{$f_2(1950)$} & 1962.8$\pm$29.3 & 132.6$\pm$22.4 & 333$\pm$61.3 & 319$\pm$42.6 &
65.4$\pm$94 & $>$119\\
{$f_2(2000)$} & 2017$\pm$21.6 & 143.5$\pm$23.3 & 614$\pm$92.6 & 58.8$\pm$24 &
450.4$\pm$221  & $>$299\\
{$f_2(2240)$} & 2207$\pm$44.8 & 136.4$\pm$32.2 & 551$\pm$149 & 375$\pm$114 &
166.8$\pm$104 & $>$222\\
{$f_2(2410)$} & 2429$\pm$31.6 & 177$\pm$47.2 & 411$\pm$196.9 & 4.5$\pm$70.8 &
460.8$\pm$209 & $>$169\\
\hline  \end{tabular}
\end{table}}
\noindent
For the background we find:\\
$\alpha_{11}=-0.0785$, $\alpha_{12}=0.0345$, $\alpha_{10}=-0.2342$,
$\beta_1=-0.06835$, \\$\gamma_1=-0.04165$,
$\beta_2=-0.981$, $\gamma_2=0.736$, $\beta_3=-0.5309$,
$\gamma_3=0.8223$.

\section{Discussion and conclusions}
\begin{itemize}
\item
In combined 3-channel model-independent analysis of data on processes
$\pi\pi\to\pi\pi,K\overline{K},\eta\eta$ in the channel with
$I^GJ^{PC}=0^+0^{++}$, an additional confirmation of the $\sigma$-meson
with mass 0.868 GeV is obtained. This mass value rather accords with
prediction ($m_\sigma\approx m_\rho$) on the basis of mended symmetry by
Weinberg \cite{Weinberg}.
In works \cite{Anisovich1}-\cite{Li-Zou-Li} evidences of the existences
of the $\sigma$-meson have been given, too.

\item
The ${f_0}(1370)$  and $f_0 (1710)$ have the dominant $s{\bar s}$ component.
Conclusion about the ${f_0}(1370)$ quite well agrees with the one of work of
Crystal Barrel Collaboration \cite{Amsler95} where the ${f_0}(1370)$ is
identified as $\eta\eta$ resonance in the $\pi^0\eta\eta$ final state of the
${\bar p}p$ annihilation at rest. Conclusion about the $f_0 (1710)$ is quite
consistent with the experimental facts that this state is observed in
$\gamma\gamma\to K_SK_S$ \cite{Braccini} and not observed in
$\gamma\gamma\to\pi^+\pi^-$ \cite{Barate}.

\item
Indication for $f_0(980)$ to be the $\eta\eta$ bound state is obtained. From
point of view of quark structure, this is the 4-quark state. Maybe, this is
consistent somehow with arguments of Refs. \cite{Achasov00,Novosib}
in favour of the 4-quark nature of $f_0(980)$.

Remembering a dispute \cite{Zou-Bugg,Morgan-Pennington} whether the $f_0(980)$
is narrow or not, we agree rather with the former. Of course, it is necessary
to make analysis of other relevant processes, first of all, $J/\psi$ and $\phi$
decays.

\item
As to the $f_0(1500)$, we suppose that it is practically the eighth component of
octet mixed with a glueball being dominant in this state. Its biggest width
among enclosing states tells also in behalf of its glueball nature
\cite{Anisovich2}.

\item
An assignment of the scalar mesons below 1.9 GeV to lower nonets is proposed.
Note that this assignment moves a number of questions and does not put the 
new ones. Now an adequate mixing scheme should be found.

\item
We do not obtain $f_2(1640)$, $f_2(1910)$, $f_2(2150)$, $f_2(2010)$, however,
we see $f_2(1450)$ and $f_2(1730)$ which are related to the 
statistically-valued experimental points.

\item
Usually one assigns to the first tensor nonet the states $f_2(1270)$ and
$f_2^{\prime}(1525)$. To the second nonet, one could assign $f_2(1601)$ and 
$f_2(1767)$ though for now the isodoublet member is not discovered. If one takes
for the isovector of this octet the state $a_2(1730)$ and if the $f_2(1601)$ is 
almost 
its eighth component, then, on the basis of the Gell-Mann--Okubo formula, we 
would expect this isodoublet mass at about 1.635 GeV. Then the relation for masses 
of nonet would be well fulfilled. Note that in the Particle Data Group issue 
\cite{PDG-04} is indicated an experiment \cite{Karnaukhov} in which one has 
observed the strange isodoublet with yet indefinite remaining quantum numbers 
and with mass $1.629\pm0.007$ GeV.   

\item
The states $f_2(1963)$ and $f_2(2207)$ together with the isodoublet $K_2^*(1980)$
could be put into the third nonet. Then in the relation for masses of nonet
\begin{equation}
M_{f_2(1963)}+M_{f_2(2207)}=2M_{K^*(1980)},
\end{equation}
the left-hand side is only 5.3 \% bigger than the right-hand one. If one consider
$f_2(1963)$ as the eighth component of octet, then the Gell-Mann--Okubo formula
\begin{equation}
M_{a_2}=4M_{K^*(1980)}-3M_{f_2(1963)}
\end{equation}
gives $M_{a_2}=2.031$ GeV. This value practically coincides with the one (2.03 GeV)
for $a_2$-meson obtained on the basis of the recent data \cite{Anis}. This state 
is interpreted \cite{Anisovich} as a second radial excitation of the 
$1^-2^{++}$-state on the basis of consideration of the $a_2$ trajectory on
the $(n,M^2)$ plane (n is the radial quantum number of the $q{\bar q}$ state).

\item
As to $f_2(2017)$, the ratio of the $\pi\pi$ and $\eta\eta$ partial widths is
in the limits obtained in Ref.\cite{Anisovich} for the tensor glueball on the
basis of the 1/N-expansion rules. However, the $K\overline{K}$ partial width is
too large for the glueball. This question requires an additional investigation.

\item
Finally we have $f_2(1450)$ and $f_2(1730)$ with the rather unusual
properties. These are non-$q{\bar q}$ states and non-glueball. Since one
predicts that masses of the lightest $q{\bar q}g$ hybrids are bigger than the ones
of lightest glueballs, maybe, these states are the 4-quark ones.

\end{itemize}

\noindent
Yu.S. and R.K. acknowledge support provided by the Bogoliubov -- Infeld Program.
M.N. were supported in part by the Slovak Scientific Grant Agency,
Grant VEGA No. 2/3105/23; and D.K., by Grant VEGA No. 2/5085/99.

\end{document}